%% file: make_astro.tex
\documentclass{mpe_report}

\usepackage{psfig,graphicx,epsfig}
\usepackage{color}
\usepackage{amsmath,amssymb,epic,eepic,array}

\begin{document}

\pagenumbering{arabic}
\setcounter{page}{173}

 \renewcommand{\FirstPageOfPaper }{173}\renewcommand{\LastPageOfPaper }{176}\include{./mpe_report_suleimanov}    \clearpage

\end{document}

%% file: mpe_report_suleimanov.tex
\newcommand{\be}{\begin{equation}}
\newcommand{\ee}{\end{equation}}

\title{Importance of Compton scattering for radiation 
spectra of isolated neutron stars}

\author{
V.~Suleimanov\inst{1,2},
K.~Werner\inst{1}}

\offprints{V.~Suleimanov}
\mail{e-mail: suleimanov@astro.uni-tuebingen.de}

\institute{
Institut f\"ur Astronomie und Astrophysik, Universit\"at T\"ubingen, Sand 1,
 72076 T\"ubingen, Germany
\and
Kazan State University, Kremlevskaja str., 18, Kazan 420008, Russia
}

\maketitle

\begin{abstract}
Model atmospheres of isolated neutron stars with low magnetic field are
 calculated with Compton  scattering taking into account.  
 Models with  effective temperatures 1, 3 and  5 MK, with two values
of surface gravity ($\log~g$ = 13.9 and 14.3), and different chemical
compositions are calculated. Radiation spectra
 computed with Compton scattering are  softer than the computed with Thomson
 scattering at high energies (E$ >$ 5 keV) for  hot ($T_{\rm eff} >$ 1
 MK) atmospheres with hydrogen-helium composition. Compton  scattering is
 more significant to hydrogen models with low surface gravity. The emergent
spectra of the hottest ($T_{\rm eff} >$ 3 MK) model atmospheres can be described
by diluted blackbody spectra with hardness factors $\sim$ 1.6 - 1.9. Compton
 scattering is less important for models with solar abundance of heavy
 elements.

\end{abstract}

\section{Introduction}

Relatively young neutron stars (NSs) with age $\le 10^6$ yr are sufficiently
hot ($T_{\rm eff} \sim 1$ MK) and  can be observed as soft X-ray
sources. Indeed, at present 
time the thermal radiation of few tens different kind of isolated NSs, from
anomalous 
X-ray pulsars to millisecond pulsars are detected. The thermal spectra of
these objects can be described by blackbody spectra with (color) temperatures
from 40 to 700 eV (see, for example, Mereghetti et. al 2002).

The plasma envelope of a NS (if it exists) can be considered 
as a NS atmosphere, and  structure and emergent spectrum of this atmosphere
can be calculated  by
using stellar 
model atmosphere methods (Mihalas 1978). Such modelling was
performed by many 
scientific groups, beginning with Romani (1987),  for NS model
atmospheres 
without magnetic field as well as for models with strong ($B > 10^{12}$ G)
magnetic field (see review by Zavlin \& Pavlov 2002). These
 model spectra were used to fit the 
observed isolated NSs X-ray spectra (see review by Pavlov et al. 2002). 

One of the important results of these works is  as follows. Emergent model spectra
of light elements (hydrogen and helium) NS atmospheres with low magnetic field
are significantly harder than
corresponding blackbody spectra. These elements are fully ionized in 
atmospheres with $T_{\rm eff} \ge$ 1 MK. Therefore, the true opacity in these
atmospheres (mainly due to free-free transitions) decreases with photon energy
as $E^{-3}$. At high energies electron scattering is larger than true
opacity and photons emitted deep in the atmosphere (where $T > T_{\rm eff}$)
escape after few scatterings on electrons. In previous works, concerning
isolated NS model atmospheres, coherent (Thomson) electron scattering is
considered. As a result, emergent spectra are very 
hard. But such a situation is very favorable to change the photon energy due
to Compton down-scattering.

It is well known that the Compton down-scattering determines the shape of
emergent model spectra of hotter NS atmospheres with $T_{\rm eff} \sim 20$
MK
 and close to Eddington limit (London et al. 1986; Lapidus et al. 1986;
Ebisuzaki 1987). These model spectra describe the
observed X-ray spectra of X-ray bursting NSs in Low-Mass X-ray Binaries
(LMXBs),  and they are close to
diluted blackbody spectra with a hardness factor $f_c \sim$ 1.5 - 1.9
(London et al. 1986; Lapidus et al. 1986, Madej 1991;  Pavlov et al. 1991).    
But these model atmospheres with Compton scattering taken into account are
not calculated for relatively cool atmospheres with $T_{\rm eff} < 10$
MK. Therefore, the effect of Compton scattering on emergent spectra of
isolated NS
model atmospheres with $T_{\rm eff} < 5$ MK is not well known up to now.   

Here we present  model atmospheres of isolated NSs with Compton
scattering taken into consideration and investigate the Compton effect on the
emergent  spectra of these atmospheres.

\section{Importance of Compton scattering}
\label{s:compton}

First of all we consider the Compton scattering effect on emergent model
spectra of isolated NS atmospheres qualitatively. It is well known that in the
non-relativistic approximation ($h\nu, kT_{\rm e} << m_{\rm e} c^2$) the
relative photon energy lost due to a scattering event on a cool electron is: 
\be
     \frac{\Delta E}{E} \approx \frac{h\nu} {m_{\rm e} c^2}.
\ee   
Each scattering event changes the relative photon energy by this value. It is
clear 
that the Compton scattering effect can be significant, if the final photon
energy change is comparable with the initial photon energy. Therefore, we
can define the Comptonisation parameter $Z_{\rm Comp}$ (see also Suleimanov et
al. 2006):
\be Z_{\rm
Compt} = \frac{h\nu}{m_{\rm e}c^2} \max((\tau_{\rm e}^*)^2,\tau_{\rm
e}^*), 
\label{e:zcomp}
\ee 
where  $\max((\tau_{\rm
e}^*)^2,\tau_{\rm e}^*)$ is the number of scattering events the photon
undergoes before escaping, $\tau_{\rm e}^*$ is the Thomson optical depth,
corresponding to the depth where escaping photons of a given frequency
are emitted. 
We can expect that Compton effects on emergent spectra of isolated NS model
atmospheres 
are significant if the Comptonisation
parameter approaches unity (Rybicki \& Lightman 1979). Because of this we
compute  $Z_{\rm Comp}$ at different photon energies (see Fig.\,1) for hot NS
model atmospheres with different chemical compositions. These models were
computed by using the method described in the next section, with the Thomson electron
scattering.  It is seen from Fig.\,1 that the Comptonisation parameter is larger (0.1
- 1) at high photon energies ($E>4 - 5$ keV) for H and He model
atmospheres. Therefore, we can expect a significant effect of Compton
scattering on the emergent spectra of these models. On the other
hand,  $Z_{\rm Comp}$ is low for the model  with solar chemical
composition of heavy elements. The Compton scattering effect has to be
weak on the emergent spectrum of this model.

\begin{figure}
\centerline{\psfig{file=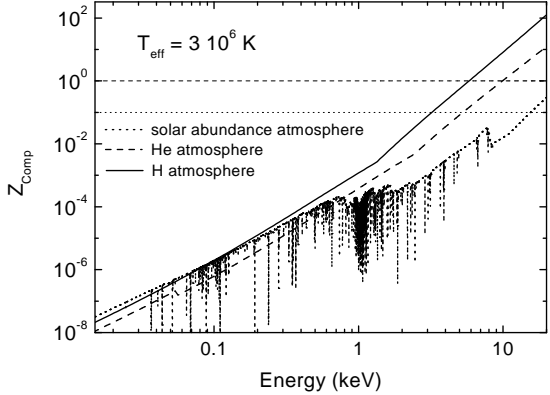,width=8.8cm,clip=} }
\caption{\label{f:fig1} 
Comptonization parameter $Z_{\rm Comp}$ vs. photon energy for neutron star
model atmospheres with different chemical composition.} 
\end{figure}
 
This qualitative analysis shows that Compton scattering can be significant for
light element model atmospheres of isolated NS we have investigated in more
detail.  

\section{Method of calculations}
\label{s:methods}

We computed model
atmospheres of isolated NSs  assuming planar geometry
using standard methods (Mihalas 1978).

The model atmosphere structure for a isolated NS with effective temperature
$T_{\rm eff}$, surface gravity $g$, and given chemical composition,  is
described by the  
hydrostatic equilibrium equation, the radiation transfer equation, and the
energy balance equation.
These equations have to be completed by the equation of state, and also
by the particle and charge conservation equations.  We assume local
thermodynamical equilibrium (LTE) in our calculations, so the number
densities of all ionisation and excitation states of all elements have
been calculated using Boltzmann and Saha equations. We take into account 
pressure ionisation effects on the atomic populations using the occupation probability
formalism (Hummer \& Mihalas 1988) as described by Hubeny et al. (1994).

Compton scattering is taken into account in 
the radiation transfer equation using the Kompaneets operator 
(Kompaneets 1957; Zavlin \& Shibanov 1991; Grebenev \& Sunyaev 2002).
The energy balance equation also accounts for Compton
scattering. 

For computing the model atmospheres we
used a version of the computer code ATLAS (Kurucz 1970; Kurucz 1993),
modified to deal with high temperatures; see Ibragimov et al. (2003)
for further details.  This code was also modified to account for Compton
scattering (Suleimanov \& Poutanen 2006; Suleimanov et al. 2006). 

Our method of calculation was tested by comparing  models for X-ray bursting neutron
star atmospheres (Pavlov et al. 1991; Madej et al. 2004), with very good
agreement. 

\section {Results}
\label{s:results}

Using this method the set of hydrogen and helium NS model atmospheres with
effective temperatures 1, 3, and 5  MK and surface gravities $\log~g$ =
13.9 and 14.3 were calculated. Models with Compton scattering and
Thomson scattering were computed for comparison. Part of the obtained results
 are presented in Figs.\,2-4.

\begin{figure}
\centerline{\psfig{file=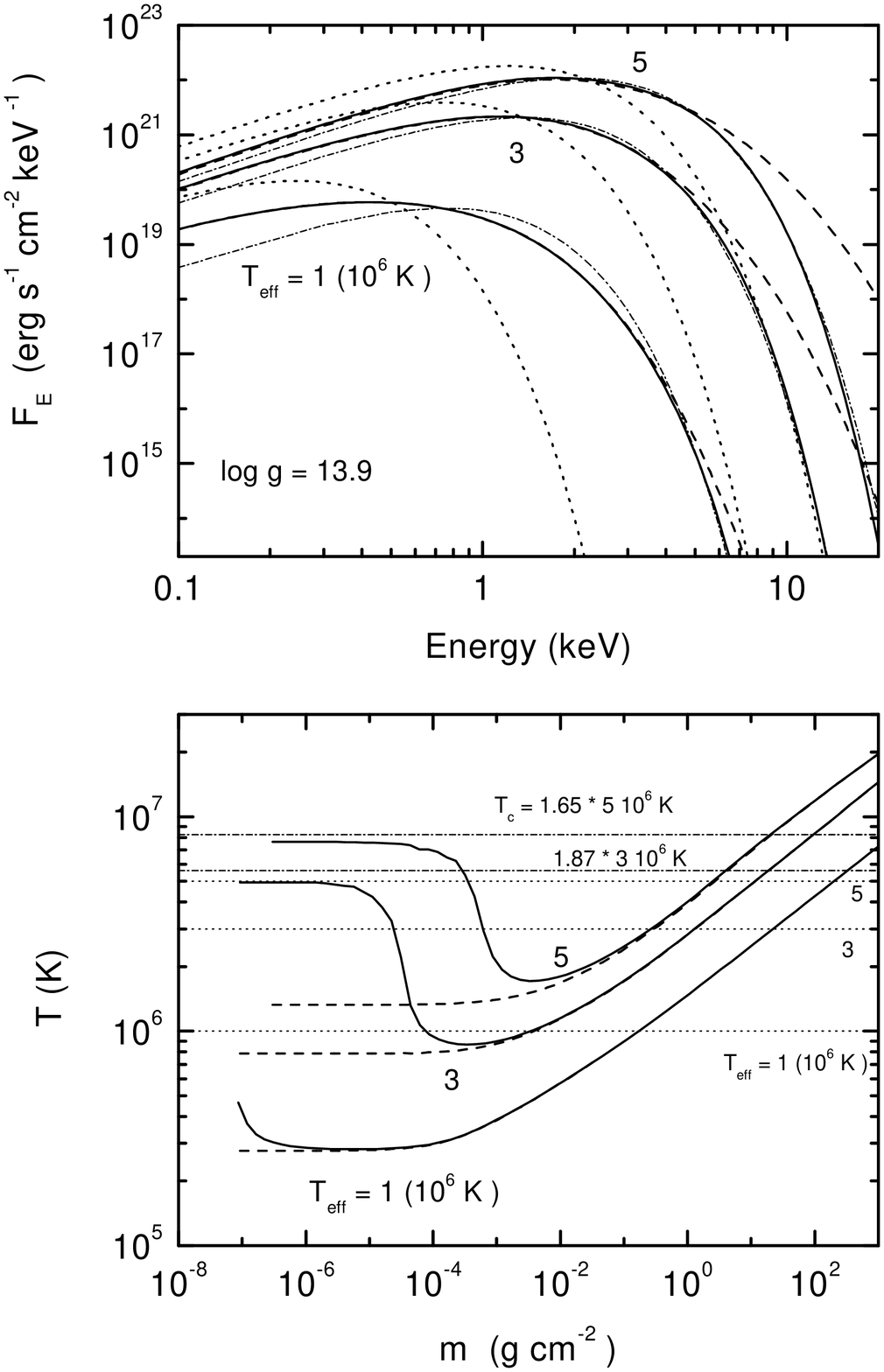,width=8.8cm,clip=} }
\caption{\label{f:fig2} 
{\it Top panel:} Emergent (unredshifted) spectra of pure H low gravity NS
model atmospheres. Solid curves - with Compton effect, dashed curves - without
Compton effect, dotted curves - blackbody spectra, thin dash-dotted curves -
diluted blackbody spectra with hardness factors 3.1, 1.87 and 1.65, for models
with $T_{\rm eff}$ = 1,3 and 5 MK. {\it Bottom panel:} Temperature structures
of the corresponding model atmospheres. Effective and color temperatures are
shown by dotted and dash-dotted lines.
}
\end{figure}

 The Compton effect is significant for spectra of hot ($T_{\rm eff} \ge 3$ MK)
hydrogen model
atmospheres at high energies (Fig.\,2). The hard emergent photons lost energy
and heat the upper layers of the atmosphere due to interactions with
electrons. As a result the high energy tails of the emergent spectra become
close to Wien spectra, and chromosphere-like structures with temperatures close
to color temperatures of the Wien spectra in the upper layers of the model
atmospheres appear. Moreover, the overall emergent model spectra of high
temperature atmospheres in first approximation can be presented as diluted
blackbody spectra with color temperatures that are close to Wien tail color
temperatures: 
\be
     F_{\rm E} = \frac{\pi}{f_{\rm c}^4} B_{\rm E} (T_{\rm c}), ~~~~~~ T_{\rm
c}= f_{\rm c} T_{\rm eff}, 
\ee   
where $f_{\rm c}$ is hardness factor. These results are similar to those obtained
for model atmospheres and emergent spectra of X-ray bursting NS in LMXBs
(see Pavlov et al. 1991; Madej et al. 2004). 

\begin{figure}
\centerline{\psfig{file=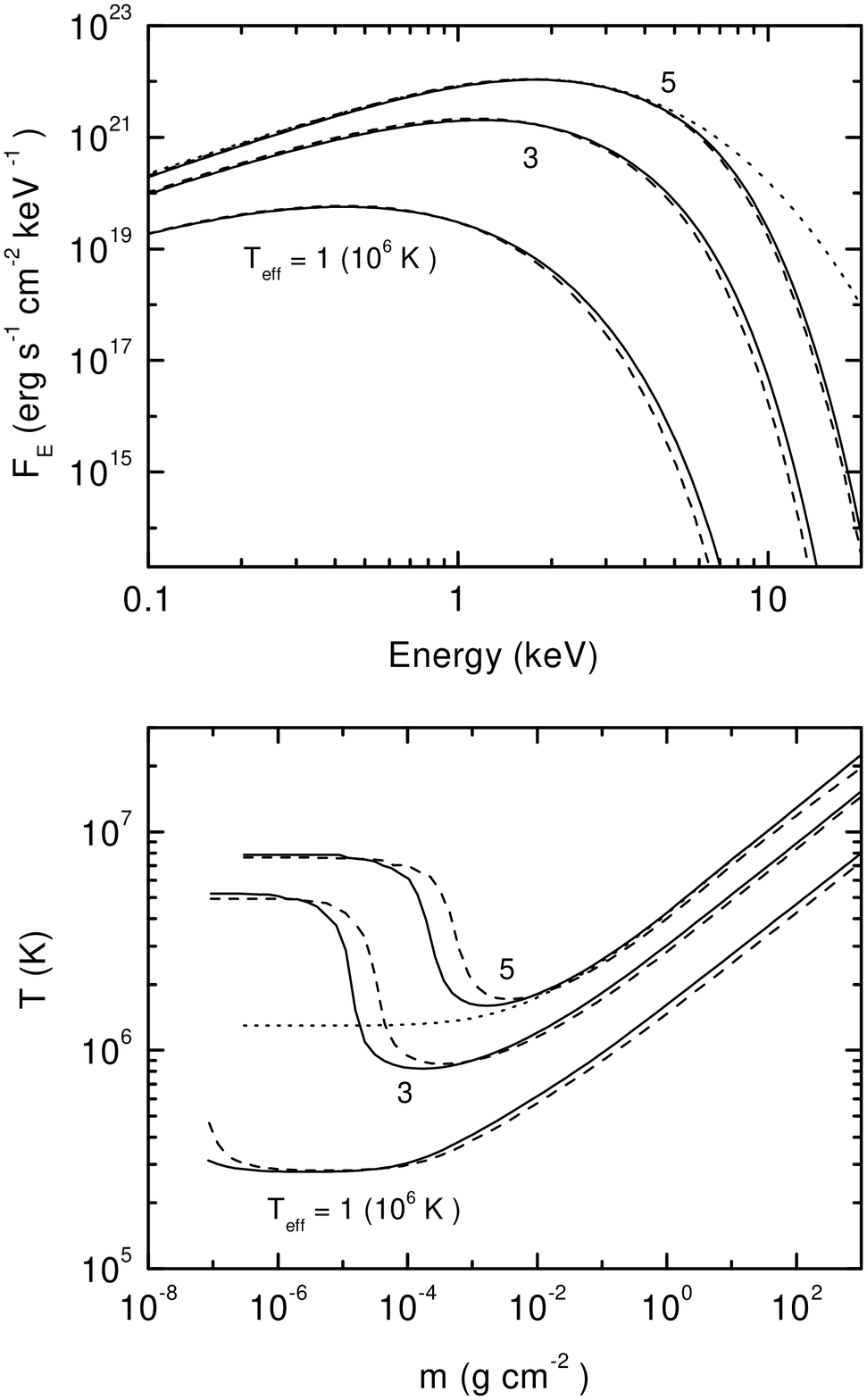,width=8.8cm,clip=} }
\caption{\label{f:fig3}
{\it Top panel:} Emergent (unredshifted) spectra of H NS model
atmospheres with different surface gravities (solid curves - high gravity
models, dashed curves - low gravity models). For comparison the model spectra
without Compton effect are shown for hottest high gravity model (dotted
curve).  {\it Bottom
panel:} Temperature structures of the corresponding model atmospheres.
}
\end{figure}

The Compton scattering effect on the emergent model spectra  of high gravity
atmospheres is less significant (Fig.\,3).
 The reason is a relatively small
contribution of electron scattering to the total opacity in high gravity
atmospheres compared to low gravity ones. 

\begin{figure}
\centerline{\psfig{file=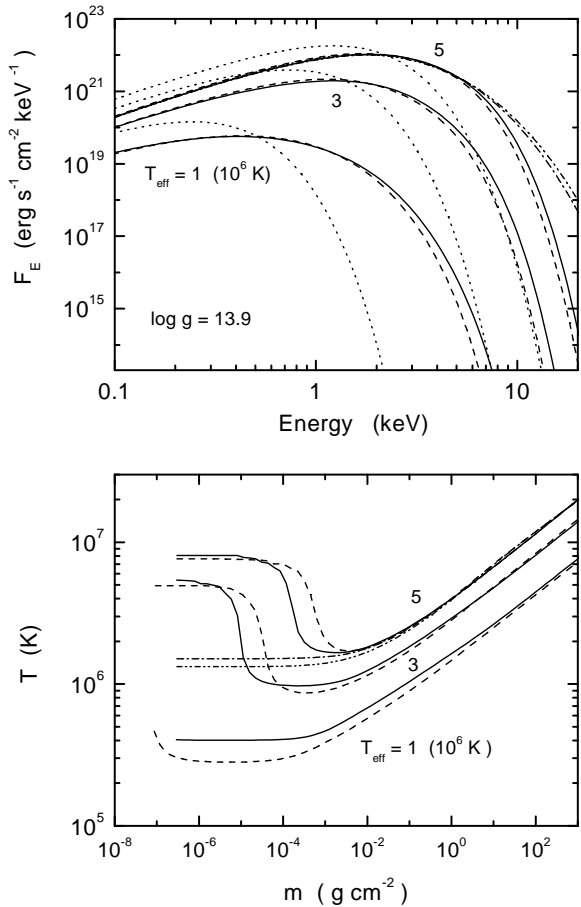,width=8.8cm,clip=} }
\caption{\label{f:fig4} 
{\it Top panel:} Emergent (unredshifted) model spectra of pure He low gravity
NS 
model atmospheres. For comparison the model spectra of pure H atmospheres are
shown by dashed curves. The model spectra of hottest atmospheres without
Compton effect are shown by dash-dotted and dash-dot-dotted curves. {\it
Bottom panel:} Temperature structures of the corresponding model atmospheres.
} 
\end{figure}

The Compton scattering effect on helium model atmospheres is also less
significant than on hydrogen model atmospheres with the same $T_{\rm eff}$ and
$\log~g$ (Fig.\,4). The reason is the same as in the case of high gravity
models. The contribution of electron scattering to the total opacity is less in the
helium models. 

\begin{figure}
\centerline{\psfig{file=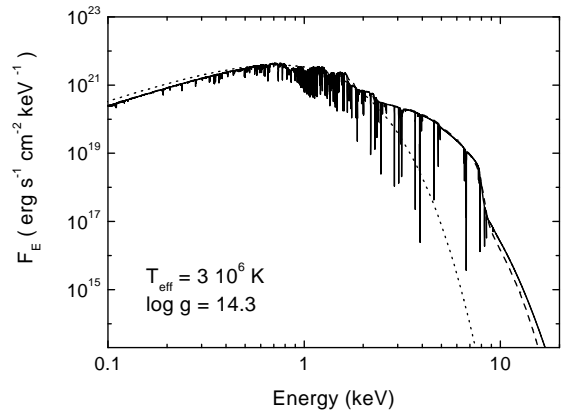,width=8.8cm,clip=} }
\caption{\label{f:fig5} 
Emergent (unredshifted) model spectra of high gravity NS atmospheres with
solar abundance of 15 most abundant heavy elements with (dashed curve) and
without (solid curve) Compton scattering. The dotted curve is the corresponding
blackbody spectrum.
} 
\end{figure}

We also computed one isolated NS model atmosphere with solar metal
abundances and  
$T_{\rm eff} = 3$ MK and $\log~g$=14.3 (see Fig.\,5). The model was
calculated with Thomson and Compton scattering and we found that the Compton effect
on the emergent spectrum is very small.

\section{Conclusions}
\label{s:conclusions}

Emergent model spectra of hydrogen and helium NS atmospheres
with $T_{\rm eff} > 1$ MK are changed by the Compton effect at high energies
($E > 5$ keV), and spectra of the hottest ($T_{\rm eff} \ge 3$ MK)
model atmospheres  can be described by diluted blackbody spectra with
hardness factors $\sim$ 1.6 - 1.9. But at the same time the spectral energy
distribution of these models are not significantly changed at the maximum of
the 
SED (at energies 1-3 keV), and effects  on the color temperatures are
not large.

The Compton effect is most significant for hydrogen
model atmospheres and to low gravity models.  Emergent model
spectra of NS atmospheres 
with solar metal abundances  are changed by Compton effects very slightly.

\begin{acknowledgements}

VS thanks DFG for financial support (grant We 1312/35-1) and Russian FBR
(grant 05-02-17744) for partial support of this investigation.

\end{acknowledgements}